\documentclass[useAMS,usenatbib,usegraphicx]{mn2e}
\usepackage{fixltx2e}

\def\aap{A\& A}

\def\apj{ApJ}
\def\apjl{ApJL}

\def\na{New Astronomy}

\def\mnras{MNRAS}

\def\mnras{{MNRAS}}

\title[Rapid Star Formation]{Rapid Star Formation and Global Gravitational Collapse}

\author[L. Hartmann et al.]{Lee Hartmann$^1$,
Javier Ballesteros-Paredes$^2$, 
Fabian Heitsch$^3$\\
$^1$ Dept. of Astronomy, University of Michigan, 500 Church St., Ann Arbor, MI 48105, USA\\
$^2$ Centro de Radioastronom\'ia y Astrof\'isica, UNAM. Apdo. Postal 72-3
(Xangari), Morelia, Michoc\'an 58089, M\'exico \\
$^3$ Dept.  of Physics and Astronomy, University of North Carolina Chapel Hill, 
CB 3255, Phillips Hall, Chapel Hill, NC 27599, USA} 

\date{MNRAS, in press }

\pagerange{\pageref{firstpage}--\pageref{lastpage}} \pubyear{2011}

\begin{document}

\label{firstpage}
\maketitle

\newcommand\msun{M_{\odot}}
\newcommand\lsun{L_{\odot}}
\newcommand\msunyr{M_{\odot}\,yr^{-1}}
\newcommand\be{\begin{equation}}
\newcommand\en{\end{equation}}
\newcommand\cm{\rm cm}
\newcommand\kms{\rm{\, km \, s^{-1}}}
\newcommand\K{\rm K}
\newcommand\etal{{et al}.\ }
\newcommand\sd{\partial}
\newcommand\mdot{\dot{M}}
\newcommand\rsun{R_{\odot}}
\newcommand\yr{\rm yr}

\begin{abstract}
Most young stars in nearby molecular clouds have estimated
ages of 1-2 Myr, suggesting that star formation is rapid.
However, small numbers of stars in these regions with 
inferred ages of $\ga 5-10$~Myr have been cited to argue 
that star formation is instead a slow, quasi-static process.  
When considering these alternative pictures
it is important to recognize that the age spread in a given star-forming cloud is
necessarily an upper limit to the timescales of local collapse,
as not all spatially-distinct regions will start contracting at 
precisely the same instant.  Moreover, star-forming clouds may 
dynamically evolve on timescales of a few Myr; in particular,
global gravitational contraction will tend to yield increasing
star formation rates with time due to generally increasing local
gas densities.  
We show that two different numerical simulations of dynamic, flow-driven 
molecular cloud formation and evolution 1) predict age spreads 
for the main stellar population roughly consistent with observations, 
and 2) raise the possibility of forming small numbers of stars early
in cloud evolution, before global contraction concentrates the
gas and the bulk of the stellar population is produced.
In general, 
the existence of a small number of older stars among a generally much-younger
population is consistent with the picture of dynamic star formation,
and may even provide clues to the time evolution of star-forming clouds.
\end{abstract}

\section{Introduction}

Roughly a decade ago, \citet[][]{BHV99} and \citet[][]{HBB01} argued
that star-forming molecular clouds in the solar neighborhood evolve
rapidly and produce stars on short - dynamical - timescales
\citep[see also][]{Elmegreen00}. The starting point for this picture 
was the observation that most nearby molecular clouds of significant mass are 
forming stars with typical ages of $\sim$ 1 - 2 Myr; only a small fraction of
the stellar population exhibits ages $\ga 5-10$~Myr.  
A straightforward interpretation of
the observations is that local star formation ensues quite quickly after
molecular cloud formation, and that lifetimes of these nearby
star-forming clouds are typically only a few Myr.  Furthermore, in
some cases the spread in ages of the bulk of the stellar population
was considerably less than a {\em lateral} crossing time. 
To explain these observations,
we proposed that molecular clouds in the solar
neighborhood tend to be formed by ``large-scale flows'', accumulating
material in a direction roughly
perpendicular to the lateral extension of the cloud 
\citep[][]{BHV99, HBB01, Heitsch_etal08b}.  Building clouds in this manner
thus eliminates the need for communicating the ``information'' needed to trigger
star formation roughly simultaneously along the length of the cloud.
The swept-up material is initially atomic; only after substantial column
densities develop, as a result of both accumulation of gas and lateral gravitational
contraction, does the cloud become molecular
\citep[][]{Bergin04}.  This evolution, driven largely by gravity 
at late stages, helps explain why star formation
is initiated shortly after ``molecular cloud formation''.\footnote{HBB01 specifically limited
their discussion to the solar neighborhood, where most of the gas is atomic and
therefore molecular clouds must be made from atomic gas.  In other regions, where
most of the gas is molecular, there should be more non-star-forming
molecular clouds.}

Since then there has been substantial discussion of apparent age
spreads in star-forming regions \citep[see][]{Jeffries11,Jeffries.etal.11}.  
One of the common findings is that, even though
the bulk of the stellar population is young, there exists a small
number of stars with apparent ages $\sim 5-10$~Myr or more
which seem to be members of the region \citep[see for example][]{Palla_etal05, 
Palla_etal07}.  The question then arises: does the
presence of a few older members in star-forming clouds contradict
the idea of dynamic cloud evolution and star formation?  Do these
apparently older stars instead indicate a long phase of quasistatic cloud 
evolution, possibly supported by turbulence and/or magnetic fields? 

In this paper we use numerical simulations to 
show that dynamic models of cloud formation can account for 
age spreads comparable to those observed, without introducing
turbulent or magnetic field support, simply because fluctuations in
initial conditions as a function of position result in some regions collapsing
faster than others.  Moreover, the dynamic models exhibit global gravitational 
collapse, which produces an increasing rate of protostellar core and star formation
over time, in a manner qualitatively similar to the
accelerating star formation rates inferred from observations by
\citet{PallaStahler99, PallaStahler00}.  A similar argument has been made
independently by \citet{AZV11}, on a semianalytical basis.

\section{A Case Study; the Orion Nebula Cluster}

To illustrate the issues typically presented by observations of young stellar
populations, we use results from recent
observational analyses of the Orion Nebula Cluster (ONC).
We focus on the ONC because it has been so well-studied, 
and because its density allows one to
limit consideration to a restricted area of the sky, thus minimizing
the problems of possible contamination (see \S 4).

\begin{figure}
\includegraphics[angle=0,width=0.5\textwidth]{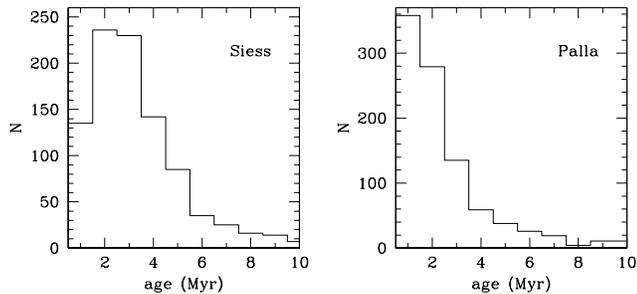}
\caption{Histograms of the estimated ages of stars in the ONC, from Da Rio \etal (2010),
using isochrones from Siess \etal (2000) (left), and Palla \& Stahler (2000, right).  The
observations, binned in linear rather than the typical log age, show a strong
skewed behavior with time (see text)}
\label{fig:histage}
\end{figure}

Figure \ref{fig:histage} shows the age distribution determined
by \citet{DaRio_etal10} using two differing sets of evolutionary
tracks, and binning linearly in age as in \citet[][]{PallaStahler00}.
(We use only the
observed members of Da Rio \etal without correction for completeness,
but the difference is negligible for our argument.)
Based on their analysis, the majority of the stars have ages $\la
4$~Myr using the \citet{Siess_etal00} evolutionary tracks, or $\la
2$~Myr using the \citet{PallaStahler99} tracks (Figure \ref{fig:histage}).
In a later study of the data, \citet[][]{Reggiani11} infer that the ONC stars are
not coeval, with star formation activity between $\sim 1.5$ and 3.5~Myr.
Similar results were found by
\citet[][]{Jeffries.etal.11}, who estimated that the apparent mean age of the ONC
is about 2.5 Myr, with 95\% of the low-mass stars formed between 1.3 and 4.8~Myr.

It is important to consider these estimates in the context of
the crossing time of the region.  The Da Rio \etal observations 
span a region of about 30 arcmin north-south, or about 3.6 pc at 
the adopted distance of 414 pc.  The (one dimensional)
velocity dispersion of stars in the region is about $2.5 - 3 \kms$
\citep{JonesWalker88, Furesz_etal08}, implying a full crossing
time of $\sim 1.2- 1.4$~Myr.  The age spreads of the main peaks
in the stellar distribution ($\sim 4$~Myr
using the Siess \etal tracks; $\sim 2$~Myr using the Palla \& Stahler
tracks) are thus $\sim 2 - 3$ crossing times. 
These values do not create a major difficulty for the picture of rapid star
formation, as the onset of gravitational collapse of individual
objects need not be coordinated better than a small number of
dynamical timescales \citep[see, e.g.,][]{Elmegreen00}.
Moreover, it should be emphasized that these dynamical timescales
refer to the {\em current} state of the region.  If, as we suggest
in \S 4, the ONC region has {\em contracted} significantly over
the last few Myr, the relevant dynamical timescale is longer than
the crossing times estimated above.

Finally, it is worth noting that observational uncertainties and problems
with theoretical isochrones can produce spurious age spreads
of similar order to those discussed above \citep[][]{Hartmann01,Hartmann03}.
Indeed, \citet[][]{Jeffries.etal.11}  suggest that this apparent
age spread is dominated by a combination of observational
uncertainties and differences in the formation processes of individual stars

The biggest challenge to dynamic models of star and cloud formation is
the ``tail'' of older stars with apparent ages between $\sim$ 5 and 10 Myr.
\citet{PallaStahler00} showed that skewed apparent age distributions
such as in Figure \ref{fig:histage} were typical of nearby star-forming regions, and
argued that this was evidence for accelerating star formation over
periods of $\sim$~10 Myr in molecular clouds.  There are observational
problems which can lead to spurious large age spreads (\S 4), but
these may not account for all of the apparently older stars,
especially objects with infrared excesses implying the
presence of circumstellar disks, or signatures of accretion
\citep[][]{Palla_etal07,Jeffries.etal.11}.
Can the dynamic picture of star formation be reconciled with the
presence of such older stars?

\section{Star formation with global gravitational collapse}

In the dynamic picture, star-forming clouds are not
in a quasi-steady state, but instead are continually evolving.
Initially, the cloud forms by sweeping up mass via large-scale flows
driven by stellar energy input or perhaps spiral density waves;
eventually, gravitational collapse leads to runaway contraction in local
regions \citep{Heitsch_etal06, Heitsch_etal08b, VSetal07,
Hennebelle_etal08, HH08, Heitsch_etal09}.  Turbulence leads to
density fluctuations which are the seeds of
subsequent gravitational collapse, modified by global contraction
which increases densities everywhere prior to disruption by outflows,
stellar winds, supernovae, etc.  Because some initial fluctuations
will create denser structures than others, some regions will collapse
before others, even without special turbulent or magnetic support; 
and because cloud densities increase with time, initially due to accumulation
of material in the post-shock regions, and later due to 
gravitational collapse, one would expect the star formation rate to 
increase with time.

To illustrate this sequence, we examine the results of two numerical
simulations of cloud formation from colliding large-scale atomic
flows.  (The mechanism is more general than the specific setup
designed for computational convenience, as any swept-up flow can be
turned into a colliding flow in a frame of reference moving with the
swept-up gas.)  It should be emphasized that these simulations
{\em minimize} differences in the onset of collapse along the cloud
because the flows collide all along the interface exactly at the
same time.  Moreover, no injected turbulent support or magnetic
fields are included. Even in these idealized cases, star formation occurs over a finite
time. 

The first simulations we consider are those of \citet{Heitsch_etal08b}.
At a resolution of $\sim 0.08$~pc, these calculations cannot follow
fragmentation down to (low-mass) protostars; instead, as a proxy for star
formation, we use the criterion of the formation of
gravitationally-bound, dense cores.  We identify cores initially via
{\tt clumpfind} \citep{Williams94}. 
We consider only gas with $T < 100$~K to avoid having to analyze all the
volume of the simulation, and demand that the local free-fall time in the clump
is at least a factor of ten smaller than the global free-fall time of the cloud.
The bulk of the core masses have temperatures near 20 K.

\begin{figure}
\includegraphics[angle=0,width=0.3\textwidth]{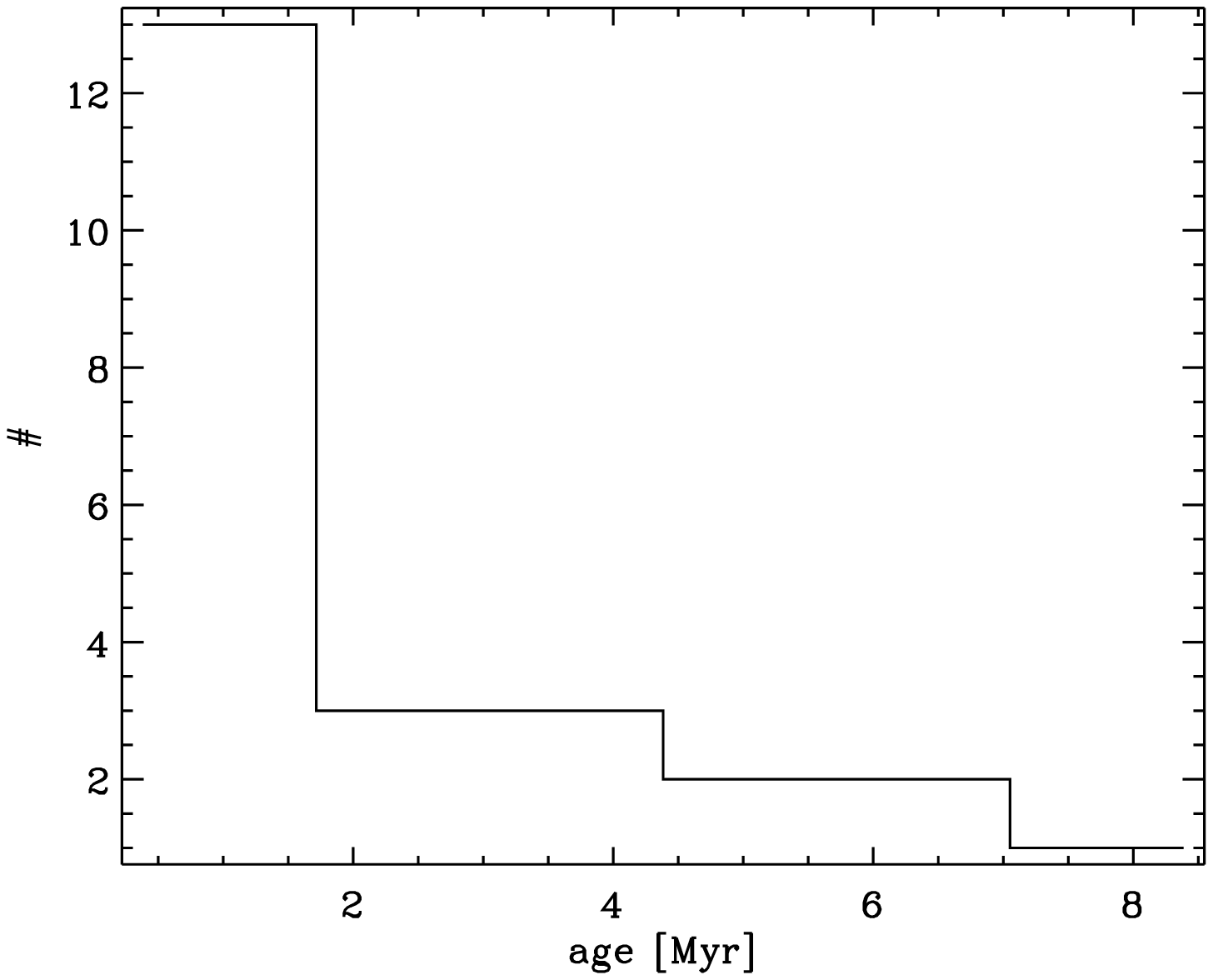} 
\includegraphics[angle=0,width=0.3\textwidth]{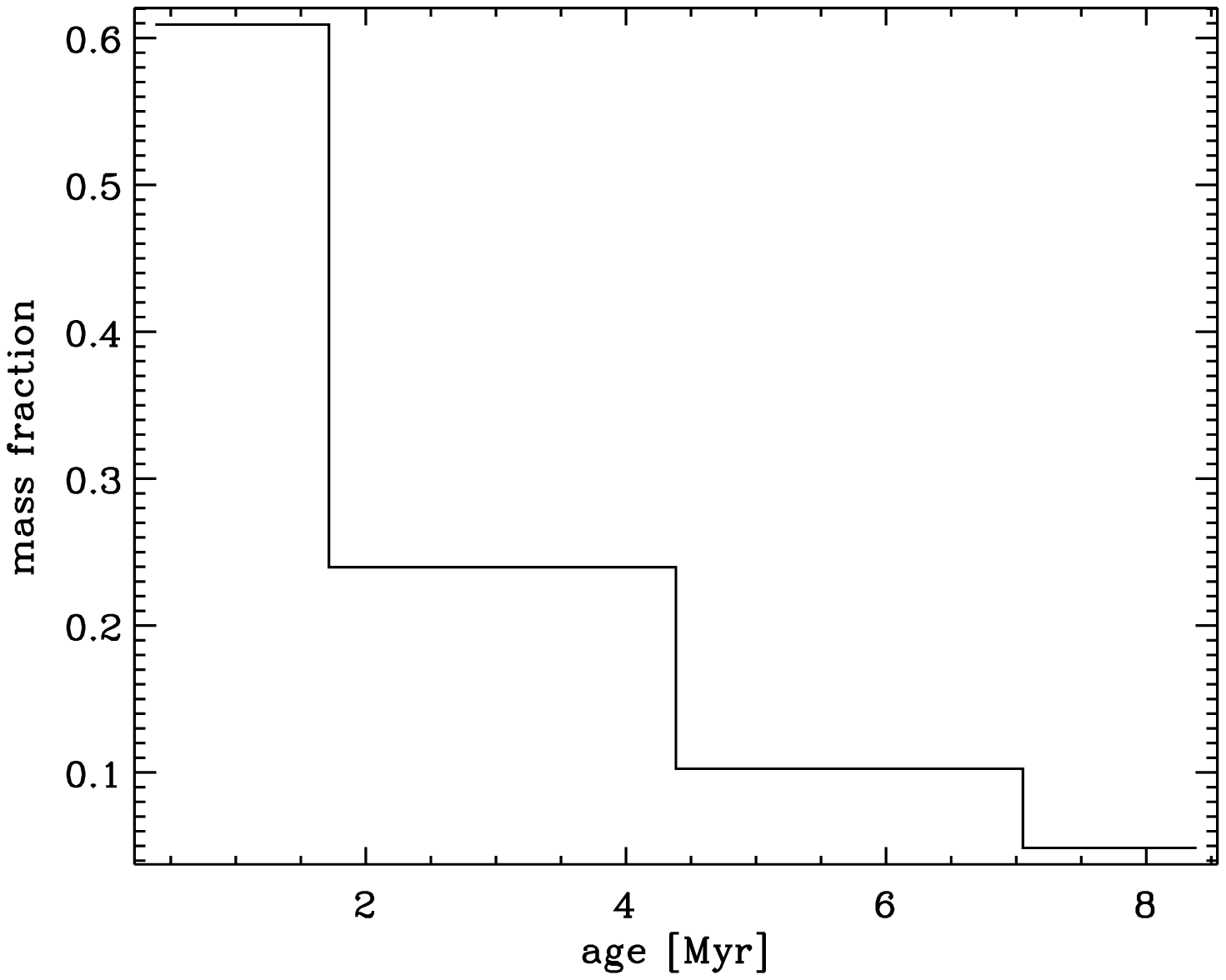}
\caption{History of massive core formation in the simulation Gs of Heitsch
\etal (2008b), plotted as a function of time prior to the end of the simulation
(see text) }
\label{fig:fhcores}
\end{figure}

Figure \ref{fig:fhcores} shows the time sequence of massive core formation in simulation
Gs of \citet{Heitsch_etal08b}.  
As these cores are quite massive, we alternatively plot both the
number of cores and the mass in cores as a function of ``look back
time'' from the end of the simulation.  Although the resolution is
limited and the timesteps are crudely binned for reasons of
statistics, it is clear that this model produces an
accelerating rate of core formation (as well as mass growth).

\begin{figure}
\includegraphics[angle=0,width=0.5\textwidth]{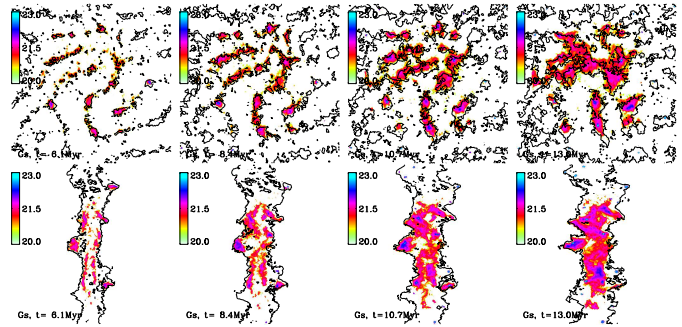}
\caption{Surface density plots for model Gs (see Figure \ref{fig:fhcores}) as
a function of time, seen face-on (upper set of panels) and edge-on
(lower set). The colors correspond to the column density in
log cm$^{_2}$ as given in the color bars.}
\label{fig:gs}
\end{figure}

Figure \ref{fig:gs} shows what is happening globally.  After a sufficient amount
of mass is accumulated in the post-shock region by the colliding flows,
gravity begins to take over, resulting in a more rapid increase in densities.
The structures are more local than in smaller cloud simulations
Gf1 and Gf2 of \citet{Heitsch_etal08b}, where the densest regions are the result of
collapse into a filament.  The latter simulations, though harder to
use for tracking core formation because less mass was involved and therefore
fewer cores were made, illustrate the type of evolution we envisage for
the ONC, where gravitational contraction over a few Myr has lead to the dense 
gas residing in the narrow ``integral-shaped filament'' (\S 4).

The cores in this simulation are quite massive, ranging from about 150 $\msun$
to about $800 \msun$.  To relate this to star formation requires an assumption
that these cores do not remain inert, but instead continue to collapse beyond
what we can determine given our resolution, fragmenting into more typical
protostellar cores of a few solar masses \citep[e.g.][]{Andre10} and then into stars. 
If the efficiency of star formation is directly related to the amount of dense core mass,
as seems reasonable \citep[see, e.g.][and references therein]{Lada10},
then the core mass and number evolution of model Gs can serve as a proxy for star formation.

The other simulations we consider are those by \citet{VSetal07},
which are comparable to the previous ones, in that two warm, thermally bistable streams 
collide to form a dense, collapsing, cold cloud. 
These simulations used {\tt gadget}, the smooth Particle Hydrodynamics (SPH)
code developed by \citet{Springel_etal01}, which includes the possibility of
sink particle formation.

\begin{figure}
\includegraphics[width=0.4\textwidth]{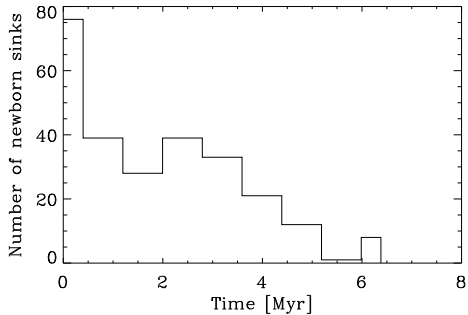}
\caption{Histograms of the newborn sink particles as a function of
time for run 20 in \citet{VSetal07}. The number of newborn sink
particles increases rapidly with time during the first 4-6 Myr.
After this time, the mass involved in sink particles is such that,
assuming a typical IMF, there would be enough massive stars to
disrupt the cloud (\citep[see for details][]{VSetal07}}.
\label{fig:vscores}
\end{figure}

In Figure \ref{fig:vscores} we show the back-in-time histogram of the
newborn sink particles for the fiducial run 20 \citep[][]{VSetal07}.
In this figure, $t=0$~Myr corresponds to $t\sim 23$~Myr since the
begining of the collision 
and thus $t=6$~Myr corresponds to the time in which the first sink
particles are formed, i.e., to $\sim$18~Myr from the
start of the simulation.\footnote{It is worth emphasizing that during
the initial evolutionary stages the flow-formed cloud is atomic, and
will only become a molecular (CO) cloud when column densities become
sufficiently large to shield the molecules from the dissociating interstellar
radiation field; see \citep[see][]{HH08}.}   From Figure \ref{fig:vscores} it is
clear that, as well as in the case of the dense cores in the previous
figure, the sink particles are being formed at an accelerated
rate.  We must note that at the final epoch plotted, the mass involved
in sink particles is such that there would be enough OB stars to
disrupt the cloud \citep[see ][]{VSetal07}, assuming a typical IMF
\citep[e.g., ][]{Kroupa01}.  Thus we omit the subsequent evolution of
the simulations as likely being unrealistic.

In both cases, the formation of dense structures (cores, sinks) occurs
at an accelerated rate due to gravitational collapse, becoming
denser and thus evolving locally faster, especially in filaments
\citep[see, e.g.,][]{Pon11}. 
While there are limitations to these simulations -
for example, there probably would be continuing fragmentation beyond
the resolution limits, and the initial conditions are
idealized - the qualitative resemblance of the simulations to the
observational results is suggestive.

\section{Discussion}

Given the current state of observational constraints, possible errors,
and uncertain physics of star formation, one must exercise caution in
interpreting the age distributions of star-forming regions 
\citep[][]{Hartmann01,Hartmann03,Jeffries11,Jeffries.etal.11}.  
We agree with \citet[][]{Jeffries.etal.11} that it is especially difficult
to be certain of age spreads in the range of $\sim 1-3$~Myr, particularly
as differing initial conditions for protostellar formation may result
in significant initial dispersions in luminosity \citep[][]{Baraffe09,
Hosokawa11,Hartmann11}.  There may even be difficulties in assigning
the ages of $\sim 10$~Myr-old stars; contamination by foreground stars 
is an issue, especially for stars without disks or accretion,
as star-forming regions are spatially-correlated. 
Even stars with disks can appear anomalously faint for their colors if 
observed edge-on, and thus mostly detected in scattered light.

Nevertheless, our simulations show that the presence of small 
numbers of older stellar members in molecular clouds 
does not pose a particular problem for the idea of 
rapid or dynamic star formation.  All such models begin with turbulent
fluctuations, and it is plausible that a few especially dense 
perturbations collapse first \citep{Heitsch_etal08b}.
The effects of global gravity then
generally results in ever-increasing densities, 
with runaway contraction in subregions, as argued
by \citet{BurkertHartmann04}, and
by \citet{HartmannBurkert07} specifically for the Orion A complex.  In
this scenario, a small number of stars are formed by a few
especially dense initial turbulent fluctuations before the overall collapse
leads to the main phase of star formation.  The two simulations presented
here suggest that time spans of 5 to 10 Myr can be accomodated
by purely dynamic models, as long as the initial star formation rate
is quite low (e.g., compare Figures 2 and 4 with Figure 1).

Absent observational problems, it is
difficult to understand the age distribution shown in Figure \ref{fig:histage}
without invoking substantial evolution of the ONC region over the last several Myr.  
For instance, if the suggestion of \citet{KrumholzMcKee05} that the star formation
rate per free-fall time is roughly a constant is correct, the age
distributions in Figure \ref{fig:histage} imply that the free-fall time has varied by
an order of magnitude, and thus the average density by two orders of magnitude,
over the last 10 Myr. 
Indeed, the recent simulations by \citet{Krumholz_etal11} of an
"ONC-like cluster" show strong evolution over $< 10^5$~yr, while a
somewhat longer contraction timescale is exhibited by the cluster
simulations of \citet{BBB03} and \citet{Bonnell_etal11}.

The suggestion of significant cloud evolution is also consistent with
kinematic studies of the ONC stars \citep{Tobin_etal09,
Proszkow_etal09}, which suggest that both the gas and stars in the ONC
are collapsing toward the central regions.  Moreover, the spatial
distribution of the stars \citep[see, for instance,][]{DaRio_etal09,
DaRio_etal10} is wider in right ascension than the narrow dense
``integral-shaped filament'' of molecular gas and
dust \citep{Bally_etal87}.  This difference is qualitatively
consistent with global gravitational collapse; many stars could have been
formed from the gas in the region in a more distended state, which has
now collapsed to form a filament, as in the simple model of
\citet{HartmannBurkert07} for Orion A.  

The idea of large-scale gravitational collapse is also consistent
with both observed column density probability density functions
\citep{BPetal11b} as well as with recent discussions showing
that the ``Larson laws'' relating velocity dispersions
(``turbulence'') with size scales are not independent of surface
density \citep{Heyer_etal09}.  These results can be interpreted as the
natural outcome of star-forming molecular clouds being in a state of
hierarchical and chaotic gravitational collapse \citep{BPetal11a}.

On the other hand, the concept of global collapse
has been challenged by Dobbs et al. (2010), who argue on
the basis of galactic-scale simulations that 
cloud-cloud collisions and stellar feedback prevent global
gravitational forces from becoming dominant.  
In a narrow sense, this is not a problem for our picture,
as we are focused on the dense star-forming regions of clouds,
which Dobbs et al. agree do become bound (and form stars).
Future observations of stellar proper motions from the Gaia spacecraft
might be able test whether or not star forming regions
are globally collapsing.

\section{Conclusions}

We have shown that models of flow-driven, dynamic dense cloud
formation and evolution predict that star formation occurs over a
finite interval of time necessarily greater than the timescales
of local collapse for individual stars.  Moreover, there is generally a strong
increase in the dense core/star formation rate over time due to the
increase in overall density and filament formation during global
gravitational collapse.  The small number of stars apparently older
than a few Myr found in or projected upon star-forming regions may be
a signature of this cloud evolution, though care must be taken to
avoid observational problems.  It may eventually be
possible to use carefully-vetted age distributions of pre-main
sequence stars to infer the global evolution of the clouds from which
they formed.

\section*{Acknowledgments}
We acknowledge a useful report from the referee.
This work was supported in part by US National Science Foundation
grant AST-0807305, by NASA grant NNX08A139G, by the University of
Michigan, and by UNAM/DGAPA grant IN110409.

\label{lastpage}

\end{document}